\documentclass[ reprint,  amsmath,amssymb,  aps]{revtex4-2}%
\usepackage{graphicx}
\usepackage{dcolumn}
\usepackage{bm}
\usepackage{amsmath}
\usepackage{amsfonts}
\usepackage{amssymb}
\usepackage{color}

\begin{document}

\title{
Removal of excess iron by annealing processes and emergence of bulk superconductivity in sulfur-substituted FeTe
}

\author{
Ryosuke Kurihara$^1$
}
\email{r.kurihara@rs.tus.ac.jp}
\author{
Ryusuke Kogure$^1$
}
\author{
Tomotaka Ota$^1$
}
\author{
Yuto Kinoshita$^2$
}
\author{
Satoshi Hakamada$^1$
}
\author{
Masashi Tokunaga$^2$
}
\author{
Hiroshi Yaguchi$^1$
}
\email{hy@rs.tus.ac.jp}

\affiliation{
$^1$Department of Physics and Astronomy, Faculty of Science and Technology, Tokyo University of Science, Noda, Chiba 278-8510, Japan	
}
\affiliation{
$^2$The Institute for Solid State Physics, The University of Tokyo, Kashiwa, Chiba 277-8581, Japan
}

\begin{abstract}
There are several strategies to discover new superconductors.
Growing new materials and applying high pressures can be the classic ways since superconductivity was found.
Also, chemical processing, such as annealing, is another way to induce superconductivity in a non-superconducting material.
Here, we show chemical processing effects in the non-superconducting material, sulfur-substituted FeTe.
It has been known that superconductivity in S-substituted FeTe is induced by O$_2$ annealing.
We revealed that hydrochloric acid etching and vacuum annealing for O$_2$-annealed samples made the quality of superconductivity higher by several physical property measurements.
Furthermore, we visualized the superconducting regions by a magneto-optical imaging technique, indicating that the superconductivity in the processed sample was bulk.
In this sample, we confirmed that the concentration of excess iron was reduced compared to that in the as-grown state.
These results provide an important route to bulk superconductivity in S-substituted FeTe and its related iron-based compounds.
\end{abstract}

\maketitle


\section{
\label{sect_intro}
Introduction
}
The discovery of high-$T_\mathrm{c}$ superconductors is attracting much attention.
The application of high-temperature superconductors is of interest to many engineers and scientists, and one of the most recent areas of concern may be the restriction on the use of superconducting magnets due to the helium crisis.
Many physicists are probably also fascinated because high-$T_\mathrm{c}$ superconductors are associated with intriguing phenomena, such as quantum criticality
\cite{Lohneysen_RMP79, Stewart_RMP83, Smidman_RMP95},
pseudo-gap
\cite{Lee_RPM78, Imajo_PRM7},
fluctuations due to non-magnetic degrees of freedom
\cite{Fernandes_PRL105, Yoshizawa_JPSJ81, Tazai_JPSJ88, Imajo_PRL125},
etc.
These may be the reasons why much effort has been devoted to searching for new superconductors since the discovery of the first superconductor Hg over 100 years ago.
Several strategies are known to discover a high-$T_\mathrm{c}$ superconductor.
One of the ways to discover a new superconductor can be to explore new materials
\cite{Steglich_PRL43, Yagubskii_JETPLett, Bednorz_ZPB64, Kamihara_JACK130}.
Another possibility is to apply high pressure to non-superconducting materials
\cite{Shimizu_JPSJ74}.
Furthermore, the chemical substitution of a non-superconducting material can also be such a way
\cite{Yi_JPSJ82, Tojo_JAP113}.
From the point of view of chemical processing, annealing can also be an important method for non-superconducting materials.

\begin{figure}[t]
\begin{center}
\includegraphics[clip, width=0.3\textwidth, bb=0 0 600 800]{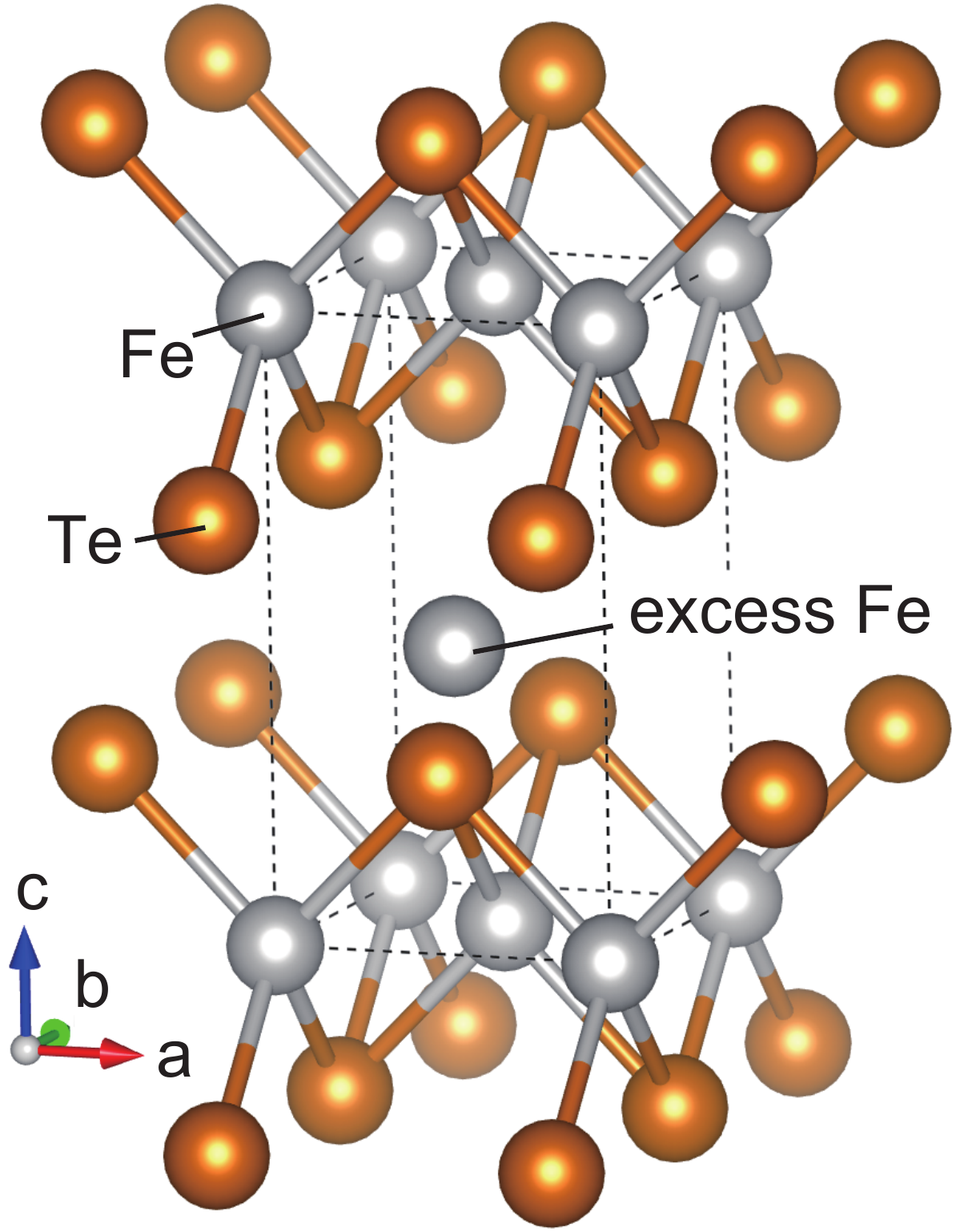}
\end{center}
\caption{
(a) Crystal structure of Fe$_{1+y}$Te with excess iron in the tetragonal phase produced by VESTA
\cite{VESTA}.
The dashed line shows the tetragonal unit cell.
The arrows with labels $a$, $b$, and $c$ indicate the crystallographic orientations.
}
\label{Photo_Samples}
\end{figure}

Figure 1(a) shows the crystal structure of the iron chalcogenide Fe$_{1+y}$Te with the space group $P4/nmm$ ($D_{4h}^7$, No. 129) at room temperature.
Here, $y$ denotes the composition of excess iron included in the crystal.
Fe$_{1+y}$Te simultaneously exhibits an antiferromagnetic (AFM) transition and a structural phase transition (SPT) from tetragonal to monoclinic with the space group $P2_1/m$ ($C_{2h}^{2}$, No. 11).
These transition temperatures range between 60 and 70 K.
The amount of excess iron also plays a key role in the AFM transition and SPT
\cite{Li_PRB79, Koz_PRB88}.
Chemical substitution by other chalcogens such as sulfur or selenium suppresses the AFM transition and SPT
\cite{Mizuguchi_JAP109, Dong_JPCM25}.
Fe-$3d$ electrons at the center of the tetrahedron formed by four Te ions mainly contribute to the formation of the Fermi surface and the antiferromagnetic character
\cite{Subedi_PRB78, Zhang_PRB82}.
Fe$_{1+y}$Te is one of the materials that does not exhibit superconductivity
\cite{Mizuguchi_PhysC469}.
Due to sulfur substitution in Fe$_{1+y}$Te, denoted as FeTe$ _{1-x}$S$_x$, superconductivity has been observed in polycrystalline samples
\cite{Ma_Vacuum195},
but several studies have reported no bulk superconductivity in single crystalline samples in their as-grown state
\cite{Mizuguchi_JAP109, Mizuguchi_IEEE21}.
On the other hand, chemical processing such as O$_2$ annealing
\cite{Mizuguchi_EPL90},
exposing to air
\cite{Mizuguchi_PRB81},
and soaking in alcoholic beverages
\cite{Deguchi_SST24}
induce superconductivity with $T_\mathrm{c} \sim 8$ K.

In the iron mono-chalcogenide systems, the amount of excess iron plays a key role in the emergence of superconductivity.
Because of the phase diagram
\cite{Okamoto_JPE12, Ipser_MC105, Mann_MTB8},
excess iron is necessary in order to obtain $\beta$-type FeSe and FeTe crystals.
Due to the excess iron, the inhomogeneous distribution of the superconducting Fe$_{1+y}$Te$_{1-x}$Se$_x$ has been observed
\cite{Okazaki_JPSJ81}.
Excess iron deintercalation has been proposed as a key role in the appearance of superconductivity
\cite{Deguchi_JAP115}.
As proposed for the Te-substituted FeSe superconductor
\cite{Sun_SciRep4},
O$_2$ annealing can remove excess iron from the sample, then superconductivity appears or a higher superconducting transition temperature is obtained.
On the other hand, according to research using samples of various sizes from a few millimeters to several tens of micrometers
\cite{Yamazaki_JPSJ85}, O$_2$ annealing for S-substituted Fe$_{1+y}$Te has been proposed to induce superconductivity only in the vicinity of the sample surface, leaving the rest of the sample non-superconducting.
Therefore, the homogeneous removal of excess iron would provide important information to induce bulk superconductivity. 

For this purpose, the combination of several chemical processes can be a promising approach. 
Previous studies on Fe$_{1+y}$Te$_{1-x}$Se$_x$ have proposed that multiple chemical processing, which consists of annealing under vacuum conditions after removing iron oxides by hydrochloric acid (HCl) etching, increases the superconducting quality
\cite{Dong_PRM3}.
We have applied this multiple processing to sulfur-substituted Fe$_{1+y}$Te compounds.
Therefore, we focus on the relationship between superconducting properties, such as the transition temperature and the superconducting shielding fraction (SSF), and the distribution of excess iron to obtain high quality superconductivity in sulfur-substituted Fe$_{1+y}$Te systems.

This paper is organized as follows.
In Sec. \ref{sect_exp}, the sample preparation and experimental procedures are described.
In Sec. \ref{Anneal}, we present the experimental results of O$_2$-annealing, HCl-etching, and vacuum-annealing effects in S-substituted Fe$_{1+y}$Te.
We demonstrate the optimized O$_2$-annealing condition in S-substituted Fe$_{1+y}$Te.
In Sec. \ref{results_MOI}, we present magneto-optical (MO) images, which serve as an experimental probe allowing direct observation of the superconducting regions
\cite{Johansen_PRB54, Kurokawa_APL116, Kinoshita_RSI93}.
In Sec. \ref{EPMA}, we show the compositional images.
These results indicate that the amount of excess iron decreases as the chemical processing progresses.
In Sec. \ref{discussion}, we discuss the contributions of excess iron to the appearance of superconductivity.
We conclude our results in Sec. \ref{conclusion}.

\section{
\label{sect_exp}
Experimental details
}

\subsection{Sample preparations and characterizations}

To grow single crystal samples, polycrystalline samples of Fe$_{1+y}$Te$_{1-x}$S$_x$ were first grown by the self-flux method
\cite{Mizuguchi_IEEE21}.
Fe shot (5N), Te shot (6N), and S shot (6N) with a nominal composition of FeTe$_{0.8}$S$_{0.2}$ were placed in a homemade quartz crucible with a diameter of 10 mm and then sealed in an evacuated quartz ampoule with a diameter of 14 mm under an atmosphere of 0.3 atm argon gas.
Single crystals of Fe$_{1+y}$Te$_{1-x}$S$_x$ were grown by Tamman's method using a homemade furnace.
The growth conditions, diagram, and crystals obtained are shown in Fig. \ref{Fig_Diagram}(a).
\cite{Yamazaki_JPSJ85, Yamamoto_JPSJ87}.
After powdering the polycrystalline sample, the powdered sample was placed in a homemade quartz crucible with a diameter of 4 mm and then sealed in an evacuated quartz ampoule with a diameter of 7 mm under an atmosphere of 0.3 atm argon gas.

\begin{figure}[htbp]
\begin{center}
\includegraphics[clip, width=0.4\textwidth, bb=0 0 400 830]{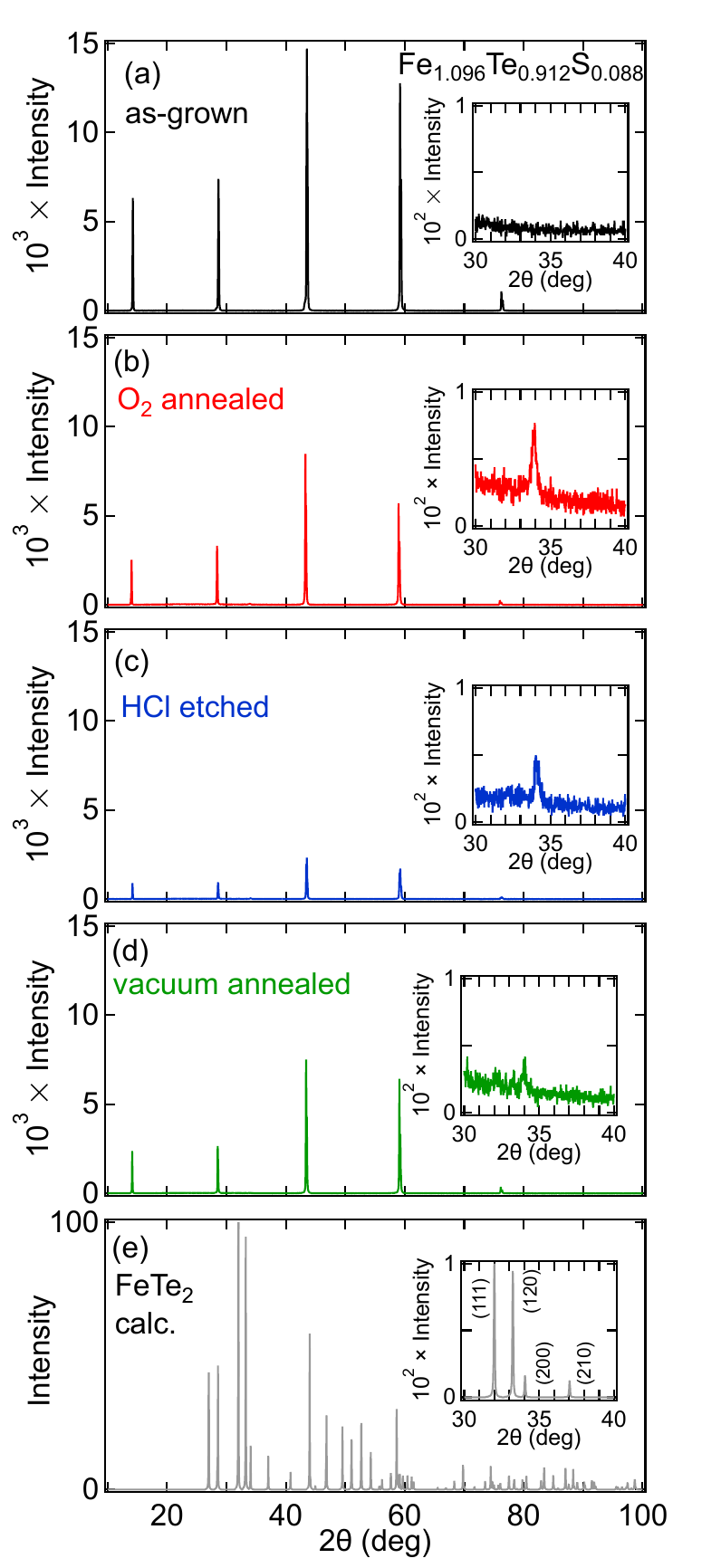}
\end{center}
\caption{
X-ray diffraction spectra of the (a) as-grown, (b) O$_2$-annealed, (c) HCl-etched, and (d) vacuum-annealed Fe$_{1+y}$Te$_{1-x}$S$_x$ with $y = 0.069$ and $x = 0.096$ and (e) calculated spectrum of FeTe$_2$.
The inset in each panel shows the x-ray diffraction spectra in the narrower region.
}
\label{Fig_XRD_FeTe2}
\end{figure}

To characterize the single crystal samples, x-ray diffraction (XRD) technique (RIGAKU, Ultima IV) was used.
We confirmed that no impurities such as iron oxide or FeTe$_2$ were contained in the as-grown samples.
The typical x-ray spectra are shown in Fig. \ref{Fig_XRD_FeTe2}.
The compositions of Fe, Te, S, and O were determined by wavelength-dispersive x-ray spectroscopy (WDS) in the Electron Probe Microanalyzer with a tungsten filament (JEOL Ltd., JXA-8100) at Research Equipment Center, Tokyo University of Science.
For the quantitative determination of the chemical compositions, FeS$_2$, Te, and SrTiO$_3$ (JEOL Ltd.) were used as standard samples.
An acceleration voltage of 15 kV and a probe current of $2 \times 10^{-8}$ A were used for the quantitative analysis.
We determined the actual atomic ratio Fe$_{1.096 \pm 0.003}$Te$_{0.912 \pm 0.001}$S$_{0.088 \pm 0.001}$ and Fe$_{1.069 \pm 0.001}$Te$_{0.904 \pm 0.001}$S$_{0.096 \pm 0.001}$ of as-grown samples. 
In the following, $y$ and $x$ represent these actual concentrations.
\begin{figure}[t]
\begin{center}
\includegraphics[clip, width=0.5\textwidth, bb=0 0 580 720]{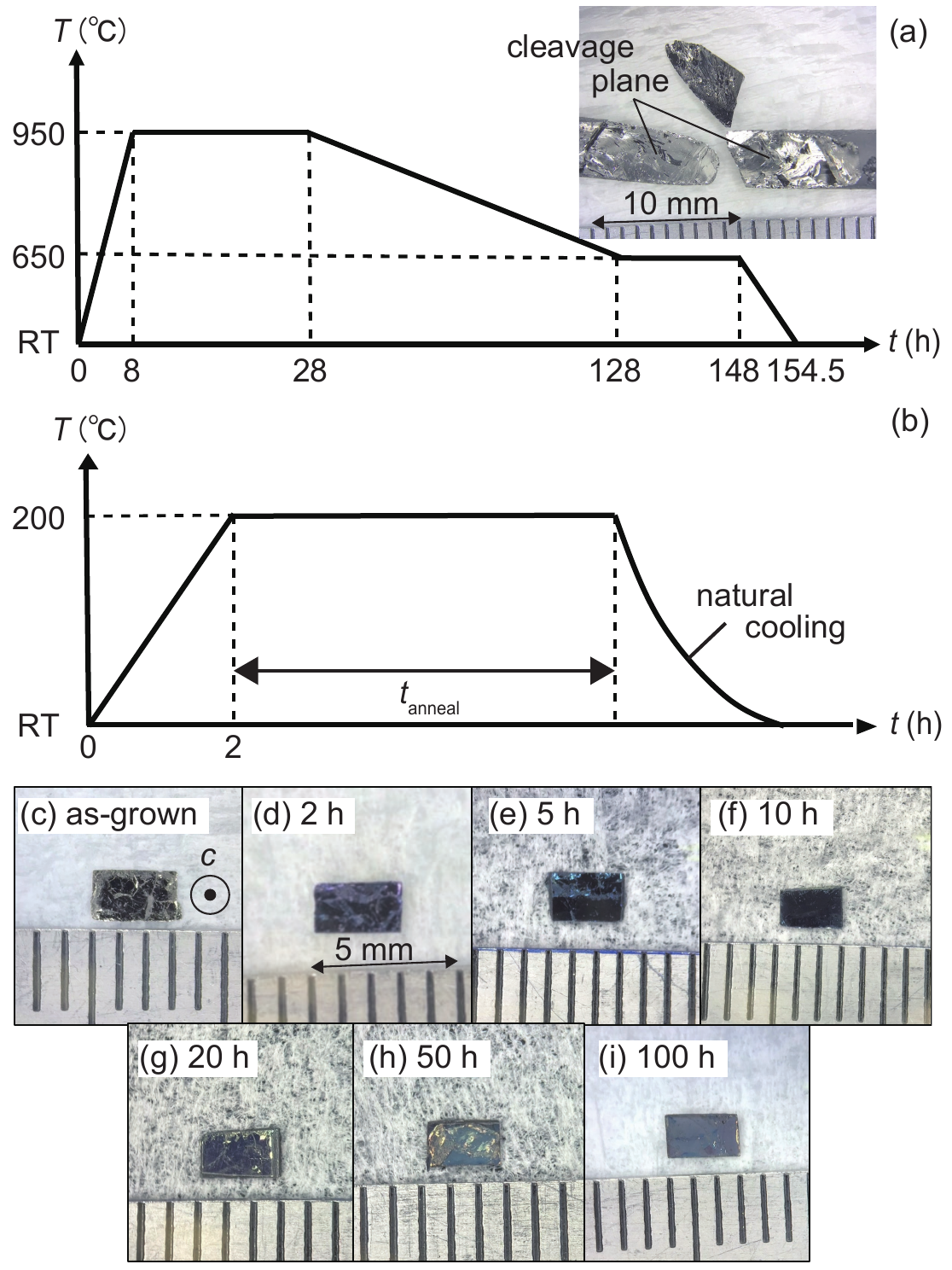}
\end{center}
\caption{
(a) Time chart of the single crystal growth of Fe$_{1+y}$Te$_{1-x}$S$_x$ by the Tamman's method.
The inset in panel (a) shiows as-grown samples.
(b) Time chart of O$_2$ annealing of Fe$_{1+y}$Te$_{1-x}$S$_x$ single crystals.
$t_\mathrm{anneal}$ indicates the O$_2$-annealing time.
RT in panels (a) and (b) is an abbreviation for room temperature. 
Photographs of O$_2$-annealed Fe$_{1.096}$Te$_{0.912}$S$_{0.088}$ with (c) $t_\mathrm{anneal} = 2$, (d) 5, (e) 10, (f) 20, (h) 50, and (i) 100 h.
}
\label{Fig_Diagram}
\end{figure}

For O$_2$ annealing, rectangular-shaped crystals were cut with a wire saw and placed on a quartz boat.
The typical dimensions of the samples were $l \times s \times t = 3 \times 2 \times 0.2$ mm$^3$.
Here, $l$, $s$, and $t$ denote the length of the long side, short side, and thickness of the sample, respectively.
This size is equivalent to the largest sample size prepared in the previous study
\cite{Yamazaki_JPSJ85}.
Atmospheric O$_2$ gas (99.99\%) was then continuously introduced into a furnace (JTEKT THERMO SYSTEMS Co., KTF035N1).
The time chart for the O$_2$ annealing is shown in Fig. \ref{Fig_Diagram}(b).
The annealing temperature of 200 $^\circ$C was chosen based on the previous study that investigated the O$_2$-annealing effect for the SSF and $T_\mathrm{c}$ at 100, 200, 300, and 400 $^\circ$C in polycrystalline FeTe$_{0.8}$S$_{0.2}$
\cite{Mizuguchi_EPL90}.
In addition, considering the sample size dependence of O$_2$ annealing
\cite{Yamazaki_JPSJ85}, 
we investigated the annealing time dependence.
After the O$_2$ annealing, the sample surface exhibited a bluish-white discoloration (see Figs. \ref{Fig_Diagram}(c) - \ref{Fig_Diagram}(i)).
Comparing the surface and interior of the annealed sample conducted by cleaving it after the O$_2$ annealing, we confirmed that the sample as a whole is not oxidized.

\begin{figure*}[htbp]
\begin{center}
\includegraphics[clip, width=1.0\textwidth, bb=0 0 730 300]{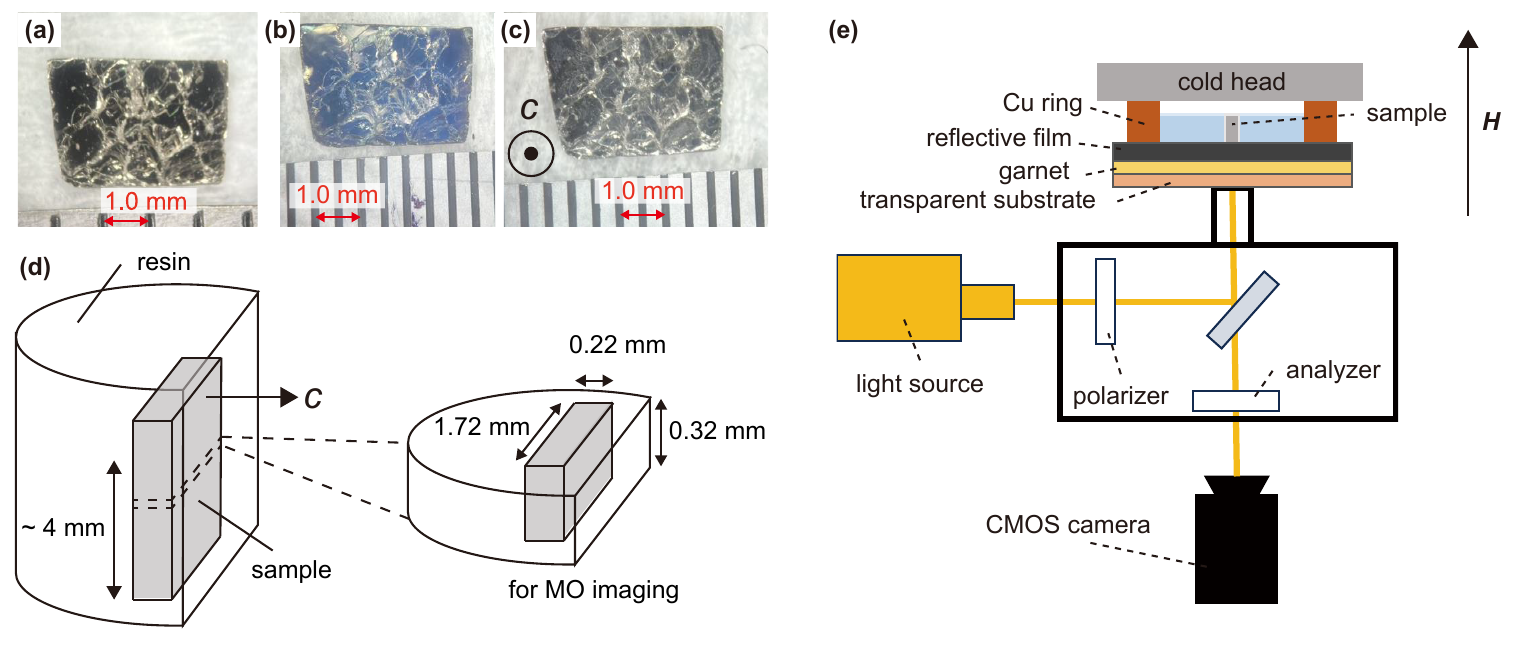}
\end{center}
\caption{
Sample photographs in (a) as-grown, (b) O$_2$-annealed, (c) HCl-etched, and (e) vacuum-annealed states.
(d) Schematic view of the sample \#MO embedded in resin (left) and cut and polished state for MO imaging measurements (right).
The right arrow in panel (d) indicates the crystallographic orientation along the $c$-axis.
(e) Schematic view of the MO measurements.
Gd$_3$Ga$_5$O$_{12}$ (GGG) is used for the transparent substrate.
}
\label{Fig_MOI}
\end{figure*}

For HCl etching to remove the iron oxides on the sample surface, room-temperature HCl with a concentration of $13.5$\% was used.
After HCl etching, the sample surface exhibited a metallic luster (see Figs. \ref{Fig_MOI}(b) and \ref{Fig_MOI}(c)).
The etching process was continued until such a surface was obtained.
The removal of the iron oxides formed on the sample surface was confirmed within the resolution of the WDS analysis.

For vacuum annealing, the quartz ampoules of $\phi$ 5 mm $\times$ 15 mm were used.
After several Ar gas flashings, the pressure was adjusted to $5 \times 10^{-3}$ Pa using a diffusion pump, and then the ampoules were sealed.
After the furnace (JTEKT THERMO SYSTEMS Co., KBF694N1) was set to the desired temperature, the sample in the quartz ampoule was immediately placed in the furnace.
After 24 hours had elapsed, the quartz ampoule was quickly quenched with ice water.
Although slightly darkened, the surface of the sample retained a metallic luster even after vacuum annealing.

\subsection{Physical property measurements}

The magnetic moment $m$ along an in-plane magnetic field direction was measured by Magnetic Property Measurement System (Quantum Design, MPMS) at the Electromagnetic Measurements Section, The Institute for Solid State Physics (ISSP), The University of Tokyo.
To estimate the residual magnetic fields of the superconducting magnet, a standard sample of Dy$_2$O$_3$ was employed.
Based on the estimated field, we measured the magnetic moment at $H = 20.0$ Oe.
The SSF was calculated from the magnetization, which is given by the magnetic moment $m$ and the sample volume $V$ as $M = m / V$, using the following relationship:
\begin{align}
\label{SSF}
\mathrm{SSF}
&= \frac{4 \pi M \times \left(-100 \right)} {H - 4 \pi M N_\mathrm{d} }
.
\end{align}
$N_\mathrm{d}$ in Eq. (\ref{SSF}) is the demagnetization factor written as
\cite{Prozorov_PRA10}
\begin{align}
N_\mathrm{d}
= \frac{4 ts}{ 4ts + 3l \left( t + s \right)}
.
\end{align}

For resistivity measurements using a current source (Lake Shore, 155) and a nanovoltmeter (HP, 34420A), the standard four-contact method was employed.
Ag paste (Fujikura, D-550) and Au wires were used to form electrodes on the sample.
A $^4$He cryostat and a homemade probe were used.

The specific heat capacity was measured by Physical Property Measurement System (Quantum Design, PPMS) at the International MegaGauss Science Laboratory, ISSP.

To investigate the effect of each chemical processing, we used a common sample in each physical property measurement.
For example, we used a single sample named \#MT for magnetic moment measurements.
We measured the magnetic moment $m$ of \#MT in its as-grown, O$_2$-annealed, HCl etched, and vacuum-annealed states.
This sequential approach can help eliminate artificial results arising from sample dependence.

\begin{table*}[htbp]
\caption{
The O$_2$-annealing time ($t_\mathrm{anneal}$) and temperature ($T_\mathrm{anneal}$), the onset of superconducting transition temperature ($T_\mathrm{c}^{onset}$), volume ($V$), demagnetization coefficient ($N_\mathrm{d}$), magnetization ($M$), and SSF for ZFC process of Fe$_{1.096}$Te$_{0.912}$S$_{0.088}$.
}
\begin{ruledtabular}
\label{table_O2 Anneal}
\begin{tabular}{ccccccc}
$t_\mathrm{anneal}$ (h)
	& $T_\mathrm{O_2}$ ($^\circ$C)
		& $T_\mathrm{c}^\mathrm{onset}$ (K)
			& $V$ ($10^{-3} $cm$^3$)
				& $10^2 \times N_\mathrm{d}$
					& $M$(1.8 K) ($10^{-1}$ emu/cm$^3$)
						& SSF(1.8 K) (\%)
\\
\hline
2
	& 200
		& 8.0
			& 1.54
				& 7.58
					& 1.31
						& 4.85
\\
5
	& 200
		& 8.0
			& 1.54
				& 8.46
					&  1.96
						& 12.9
\\
10
	& 200
		& 8.1
			& 1.42
				& 9.89
					&  2.62
						& 17.1
\\
20
	& 200
		& 8.4
			& 1.40
				& 9.33
					&  5.85
						& 36.4
\\
50
	& 200
		& 8.6
			& 0.82
				& 9.06
					&  5.55
						& 34.7
\\
100
	& 200
		& 8.6
			& 1.48
				& 9.11
					&  3.68
						& 30.2
\end{tabular}
\end{ruledtabular}
\end{table*}

\subsection{Magneto-optical imaging}

For MO imaging measurements, we used one sample named \#MO.
Figures \ref{Fig_MOI}(a) - \ref{Fig_MOI}(c) present photographs of the as-grown, O$_2$-annealed, and HCl etched \#MO.
The sample \#MO was cut and polished after being embedded in epoxy resin (Henkel Japan Ltd., STYCAST$^\mathrm{TM}$ 1266J) (see Fig. \ref{Fig_MOI}(d)).

Figure \ref{Fig_MOI}(e) exhibits the schematic view of the MO measurements.
The MO images were obtained by a polarized light microscope (OLYMPUS, BXFM) with a 100-W halogen lamp (OLYMPUS, U-LH100L-3) at the International MegaGauss Science Laboratory, ISSP.
In the experiment, the analyzer was shifted from orthogonal to the polarizer by 2$^\circ$.
An imaging sensor with the reflective film, magnetic active garnet film, and transparent substrate was employed.
A $^4$He cryostat was used to cool the sample well below the superconducting transition temperature.
The imaging sensor with the embedded \#MO attached was mounted onto the cold head of the cryostat via a copper ring.
To obtain clear images, we integrated $100$ captured images.
For generating magnetic fields up to 12.6 Oe, a homemade Helmholtz coil with a diameter and a gap of 100 mm and a current source were used.
In the range of weak magnetic fields, the light intensity of the image is proportional to the magnetic field.
Thus, we can illustrate the spatially-resolved magnetic field distribution.
\cite{Goa_RSI74}.

\section{
\label{sect_3}
Experimental results
}

\subsection{
\label{Anneal}
Annealing and etching effects on superconductivity
}

In this section, we present the results of the chemical processing and their impact on the superconducting properties, including the transition temperature and superconducting shielding fraction.
We find that the vacuum annealing following the HCl etching process for the O$_2$-annealed samples leads to an improvement in these properties.

\begin{figure}[t]
\begin{center}
\includegraphics[clip, width=0.5\textwidth, bb=0 0 600 280]{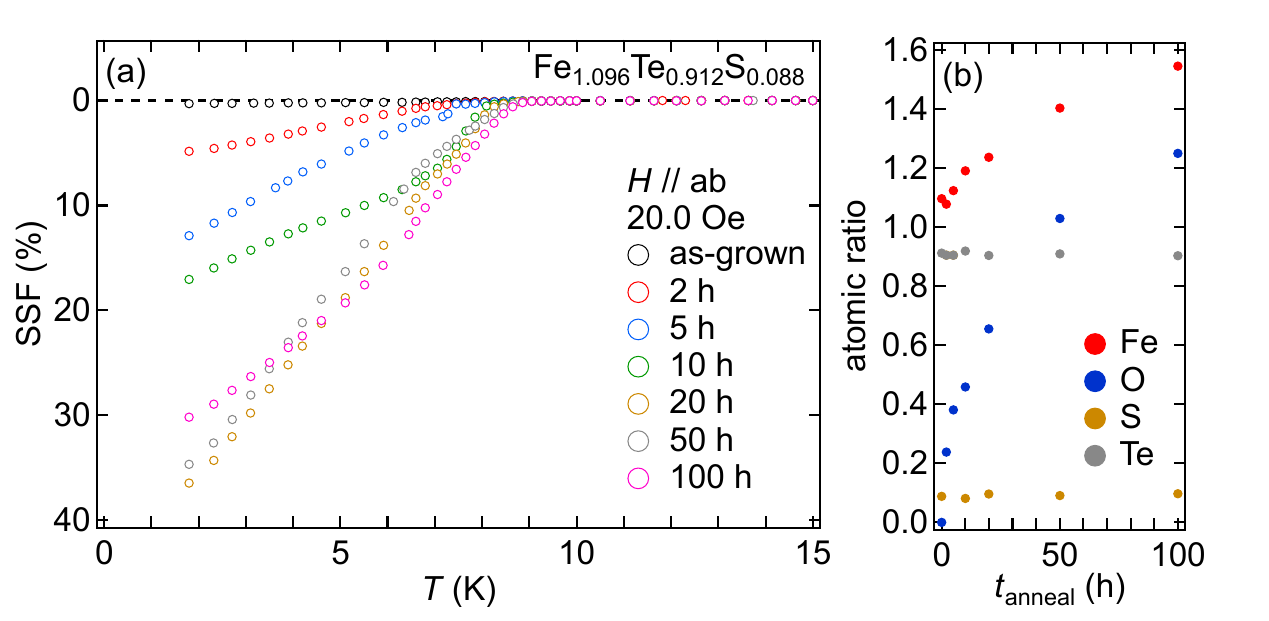}
\end{center}
\caption{
(a) Temperature dependence of the SSF of several O$_2$-annealed Fe$_{1+y}$Te$_{1-x}$S$_x$ with $y = 0.096$ and $x = 0.088$ for ZFC process.
The annealing temperature for each sample is set to 200 $^\circ$C.
A magnetic field of 20.0 Oe was applied along the $ab$-plane.
(b) O$_2$-annealing time dependence of the atomic ratio on the $(001)$ surface of the samples determined by WDS.
Each atomic ratio is normalized by the atom\% of Te + S.
The uncertainty of the atomic ratio of Fe, Te, and S is estimated to be $0.005$, and that of O is estimated to be 0.05.
}
\label{MT_O2_Anneal}
\end{figure}

\begin{table*}[t]
\caption{
Information of each chemical processed sample in Fe$_{1+y}$Te$_{1-x}$S$_x$.
The values of SSF for the ZFC process are also listed.
}
\begin{ruledtabular}
\label{table_ProcessDep_1}
\begin{tabular}{cccccccc}
sample
    &composition
	   &process
		& $t_\mathrm{process}$
			& $T_\mathrm{c}^\mathrm{onset}$ (K)
				& $V$ ($10^{-3} $cm$^3$)
					& $10^2 \times N_\mathrm{d}$
						& SSF(1.8 K) (\%)
\\
\hline
\#MT1
    &$1 + y = 1.096 \pm 0.003$ 
	   & O$_2$
		& 50 h
			& 8.6
				& 0.82
					& 9.06
						& 33.8
\\
    &$x = 0.088 \pm 0.001$
	   &HCl
		& 100 min
			& 8.1
				& 1.71
					& 7.22
						& 34.6
\\
    &
	   &vacuum
		& 24 h
			& 8.5
				& 0.93
					& 9.79
						& 68.1
\\
\#MT2
    &$1 + y = 1.069 \pm 0.001$
	   & O$_2$
		& 50 h
			& 8.4
				& 1.10
					& 6.52
						& 39.1
\\
    &$x = 0.096 \pm 0.001$
	   &HCl
		& 90 min
			& 8.6
				& 1.92
					& 8.53
						& 47.9
\\
    &
	   &vacuum
		& 24 h
			& 8.6
				& 1.92
					& 8.53
						& 57.6
\end{tabular}
\end{ruledtabular}
\end{table*}

First, we demonstrate the investigation of the optimal condition of O$_2$ annealing for Fe$_{1+y}$Te$_{1-x}$S$_x$ with $y = 0.096$ and $x = 0.088$.
Figure \ref{MT_O2_Anneal}(a) shows the temperature dependence of the SSF for the zero field cool (ZFC) process calculated from the magnetic moment of several O$_2$-annealed samples.
Here, the O$_2$ annealing was applied to each of the different as-grown samples.
We confirmed that the as-grown sample did not show superconductivity down to 1.8 K.
In contrast, the sample O$_2$ annealed for 2 hours shows diamagnetic signals below 7 K, indicating that the O$_2$ annealing induces superconductivity.
The highest $T_\mathrm{c}^\mathrm{onset} = 8.6$ K is obtained by 50- and 100-hour annealed sample (see Table. \ref{table_O2 Anneal}).
On the other hand, the 20- and 50-hour annealed samples show 36.4\% and 34.7\% SSF, respectively.
Thus, we concluded that the optimal condition for 200 $^\circ$C O$_2$ annealing is obtained with the aformentioned annealing duration.
We note that the diamagnetic signals gradually increase below $T_\mathrm{c}$, suggesting that the superconductivity in the O$_2$-annealed samples is inhomogeneous or impurities are included in the samples.

To confirm the oxidation, we determined the actual atomic ratio of the sample surface by WDS.
Figure \ref{MT_O2_Anneal}(b) shows the O$_2$-annealing time dependence of the atomic ratio determined by WDS.
Here, these ratios are normalized to that of Te + S. 
As the annealing time is increased, the atomic ratio of Fe and O on the $(001)$ sample surface increases from the as-grown value.
This result indicates that iron oxides such as Fe$_2$O$_3$ and Fe$_3$O$_4$ are formed on the surface.

\begin{figure}[t]
\begin{center}
\includegraphics[clip, width=0.5\textwidth, bb=0 0 440 520]{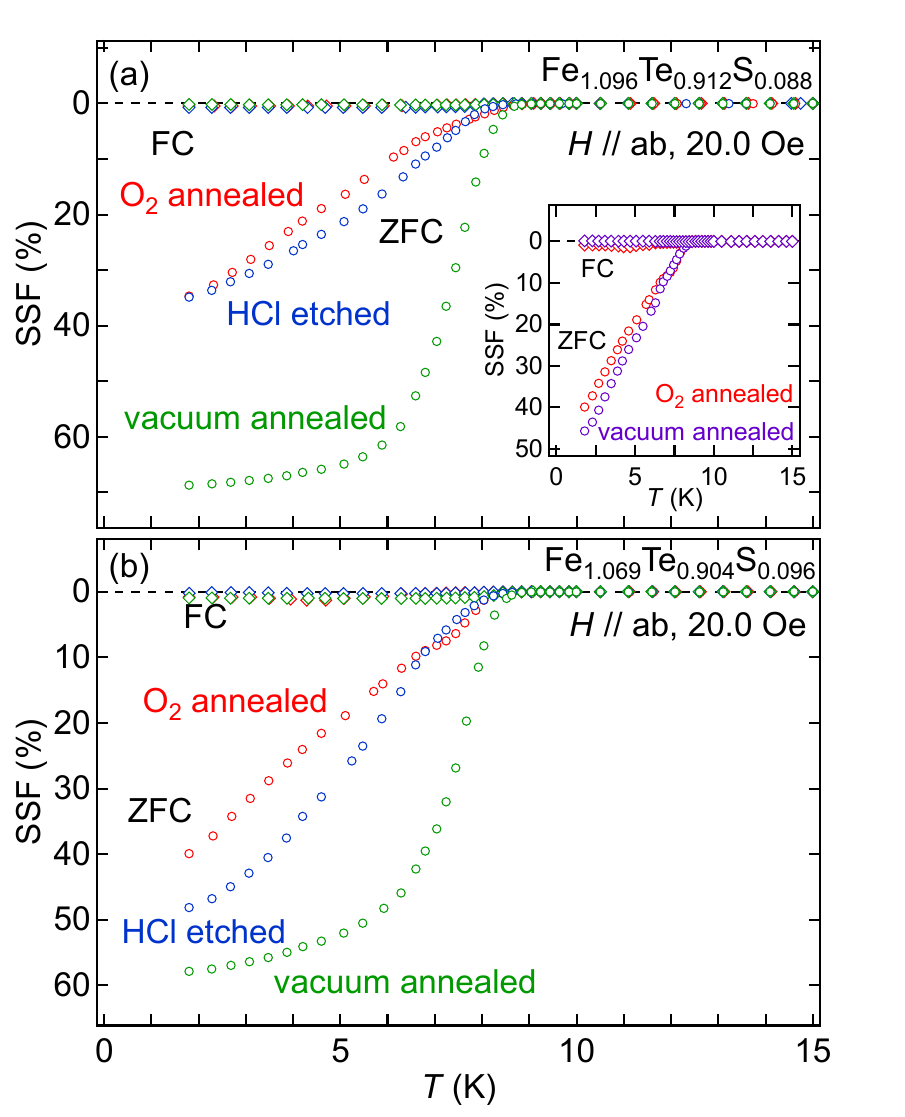}
\end{center}
\caption{
(a) Temperature dependence of the SSF of the O$_2$-annealed, HCl-etched, and vacuum-annealed Fe$_{1+y}$Te$_{1-x}$S$_x$ with $y = 0.096$ and $x = 0.088$ (\#MT1).
A magnetic field of 20.0 Oe was applied along the ab plane.
The inset in panel (a) shows the SSF of the O$_2$-annealed sample and the vacuum-annealed sample without HCl etching.
(b) Temperature dependence of the SSF for the ZFC process of the HCl etched and vacuum-annealed Fe$_{1+y}$Te$_{1-x}$S$_x$ with $y = 0.069$ and $x = 0.096$ (\#MT2).
A magnetic field of 20.0 Oe was applied along the $ab$-plane.
}
\label{AnnealProcessDep}
\end{figure}

After the O$_2$ annealing, we performed HCl etching.
Figure \ref{AnnealProcessDep}(a) shows the temperature dependence of the SSF of the processed Fe$_{1+y}$Te$_{1-x}$S$_x$ with $y = 0.096$ and $x = 0.088$ (\#MT1).
Here, \#MT1 was annealed in O$_2$ at 200 $^\circ$C and for 50 hours.
The magnetization exhibits diamagnetic signals, indicating that the HCl etching hardly removes the superconducting regions of samples.
Comparing the SSF before and after the processing in Fig. \ref{AnnealProcessDep} and Table \ref{table_ProcessDep_1}, we deduce that the effect of the HCl etching is not very significant for superconductivity.
We also determined the composition Fe$_{0.949}$Te$_{0.912}$S$_{0.088}$ of the HCl-etched sample by WDS, indicating that the iron oxides on the surface are removed.
This result indicates that excess iron is removed from the surface of the sample, or FeTe$_2$ (Fe$_2$) is formed on the surface.

In contrast to the HCl etching, a drastic change in the superconducting properties was observed after the vacuum annealing.
Comparing the SSF of the vacuum-annealed \#MT1 with that of the HCl-etched one (see Fig. \ref{AnnealProcessDep}(a) and Table \ref{table_ProcessDep_1}), we can see that the magnetization of the vacuum-annealed \#MT1 decreases rapidly below $T_\mathrm{c}$.
This result indicates that the vacuum annealing after the HCl etching improves the quality of superconductivity.
In addition, we confirmed the effectiveness of the vacuum annealing on different Fe$_{1+y}$Te$_{1-x}$S$_x$ with $y = 0.069$ and $x = 0.096$ (\#MT2) as well (see Fig. \ref{AnnealProcessDep}(b)).
Therefore, we conclude that vacuum annealing after the HCl etching is an effective method to obtain superconductivity.

The importance of the HCl etching process is also confirmed by the vacuum annealing effect of Fe$_{1+y}$Te$_{1-x}$S$_x$ without HCl etching (see the inset in Fig. \ref{AnnealProcessDep}(a)).
Here, the vacuum-annealed sample is identical to the O$_2$-annealed one.
The SSF of the vacuum-annealed samples without the HCl-etching process is similar to that of the O$_2$-annealed sample.
Therefore, we can conclude that the HCl-etching process before the vacuum-annealing process enhances the superconducting properties.

\begin{table}[t]
\caption{
Twenty-point average of atom\% of Fe, Te, S, and O and its standard deviation (SD) in the vacuum-annealed Fe$_{1.096}$Te$_{0.912}$S$_{0.088}$ near the sample surface parallel to the $ab$-plane.
Each value was determined by WDS.
These normalized values by Te + S and their values of uncertainty (VU) are also listed.
}
\begin{ruledtabular}
\label{table_FeTe2}
\begin{tabular}{ccccc}
&Fe
	&Te
		&S
			&O
\\
\hline
atom\%
&45.2 
	& 45.4   
		& 4.69   
			& 4.7  
\\
SD
&1.0 
	&1.8   
		&0.35   
			&2.1  
\\
normalized
&0.903 
	&0.906    
		&0.094    
			&0.095   
\\
VU
&0.019 
	&0.003    
		&0.003    
			&0.036   
\\   
\end{tabular}
\end{ruledtabular}
\end{table}

Based on the previous study in Ref. 34
,
we attempted a second round of chemical processing for the vacuum-annealed sample.
We obtained an O$_2$-annealed sample, but we were unable to obtain a sample with a metallic luster surface after the HCl etching process, whereas the previous study performed multiple annealing processes.
This result can be attributed to the formation of FeTe$_2$ on the sample surface by the optimized O$_2$-annealing condition.
As shown in Fig. \ref{Fig_XRD_FeTe2}, XRD measurements indicate the formation of FeTe$_2$ because we observed a peak structure around $34^\circ$ except for the as-grown sample.
We deduce that this peak is attributed to the $\left( 200 \right)$ peak of FeTe$_2$.
Due to insufficient reorientation of FeTe$_2$, other peaks may not be visible.
Compositional analysis near the sample surface also indicates the contributions of FeTe$_2$ and FeS$_2$ in addition to Fe$_{1+y}$Te$_{1-x}$S$_x$.
Table \ref{table_FeTe2} shows the WDS results of another vacuum-annealed Fe$_{1+y}$Te$_{1-x}$S$_x$ with $y = 0.096$ and $x = 0.088$.
In this sample, we were unable to remove iron oxides because of the residual oxygen.
Based on the normalized values by the atom\% of Te + S, we consider that the amount of Fe atoms is significantly lower compared to the composition of the as-grown sample.
Assuming that the surface is composed of Fe$_{1.096}$Te$_{0.912}$S$_{0.088}$, FeTe$_2$, FeS$_2$, and Fe$_2$O$_3$, we estimate their respective proportions as shown in Table. \ref{table_FeTe2_2}, indicating the formation of FeTe$_2$ and FeS$_2$ on the Fe$_{1+y}$Te$_{1-x}$S$_x$ surface by the annealing process.
Further detailed studies are required regarding the formation of FeTe$_2$ on Fe$_{1+y}$Te$_{1-x}$S$_x$ and its related compounds.
\begin{table}[t]
\caption{
Formula percent that consists of the sample near the surface of the vacuum-annealed Fe$_{1.096}$Te$_{0.912}$S$_{0.088}$.
Each value is estimated from the compositions listed in Table \ref{table_FeTe2}.
}
\begin{ruledtabular}
\label{table_FeTe2_2}
\begin{tabular}{cccc}
Fe$_{1+y}$Te$_{1-x}$S$_x$ ($y = 0.096$, $x = 0.088$)
	&FeTe$_2$
		&FeS$_2$
			&Fe$_2$O$_3$
\\
\hline
0.70    
	&0.24  
		&0.03 
			&0.04    
\end{tabular}
\end{ruledtabular}
\end{table}

\begin{figure}[t]
\begin{center}
\includegraphics[clip, width=0.5\textwidth, bb=0 0 430 300]{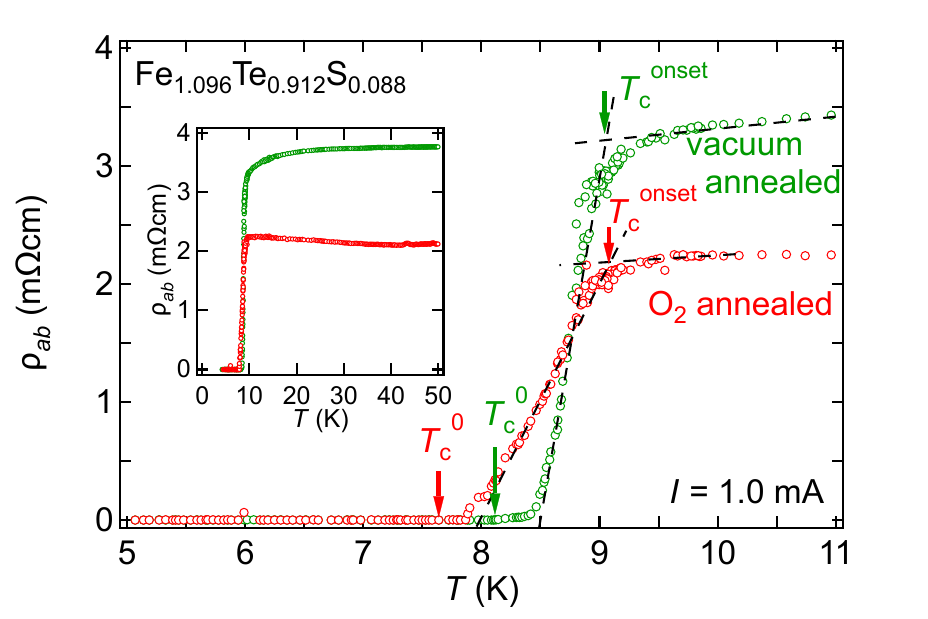}
\end{center}
\caption{
Temperature dependence of the in-plane electrical resistivity $\rho_\mathrm{ab}$ of the O$_2$- and vacuum-annealed Fe$_{1.096}$Te$_{0.912}$S$_{0.088}$ (\#RT).
The dashed lines indicate the linear fit of $\rho_\mathrm{ab}$.
$T_\mathrm{c}^0$ is determined by fitting the slope of $\rho_{ab}-T$.
The inset shows the electrical resistivity below 30 K.
}
\label{Fig_rhoT}
\end{figure}

We also investigated the effectiveness of vacuum annealing on Fe$_{1+y}$Te$_{1-x}$S$_x$ in terms of electric transport properties.
Figure \ref{Fig_rhoT} shows the temperature dependence of the electrical resistivity in the O$_2$-annealed Fe$_{1+y}$Te$_{1-x}$S$_x$ with $y = 0.096$ and $x = 0.088$  (\#RT).
$\rho_{ab}$ shows a resistive drop at temperatures below the onset of the superconducting transition $T_\mathrm{c}^\mathrm{onset}$.
For further decreasing temperatures, $\rho_{ab}$ becomes zero at $T_\mathrm{c}^0$.
As listed in Table \ref{table_RT}, we determined $T_\mathrm{c}^0$ and $T_\mathrm{c}^\mathrm{onset}$ to be  $7.6$ and $9.1$ K, respectively.
Using these values, we obtained the relative variation of the superconducting transition temperature, given by
$\mathit{\Delta}T_\mathrm{c}/T_\mathrm{c} = \left( T_\mathrm{c}^\mathrm{onset} - T_\mathrm{c}^0 \right)/ T_\mathrm{c}^\mathrm{onset}$, to be 0.16.
To perform the vacuum annealing after the HCl etching on the O$_2$-annealed \#RT, $T_\mathrm{c}^0$ and $T_\mathrm{c}^\mathrm{onset}$ change to 8.1 and 9.0 K, respectively.
Due to these increases in the characteristic temperatures, the relative variation $\mathit{\Delta}T_\mathrm{c}/T_\mathrm{c}$ is reduced to 0.10.
Because the relative variation of the superconducting transition temperature reflects the quality of superconductivity
\cite{Wu_PhysC469, Lei_MCP127}, we deduce that the narrowing of the transition temperature width is attributed to the reduction of defects or impurities by removing excess iron.

\begin{table}[t]
\caption{
$T_\mathrm{c}^0$, $T_\mathrm{c}^\mathrm{onset}$, and $\mathit{\Delta}T_\mathrm{c}/T_\mathrm{c}$ of Fe$_{1.096}$Te$_{0.912}$S$_{0.088}$ (\#RT) determined by the resistivity rate measurements.
}
\begin{ruledtabular}
\label{table_RT}
\begin{tabular}{cccc}
process
    &$T_\mathrm{c}^0$ (K)
	   &$T_\mathrm{c}^\mathrm{onset}$ (K)
        &$\mathit{\Delta}T_\mathrm{c}/T_\mathrm{c}$
\\
\hline
O$_2$
	&7.6
		&9.1
			&0.16
\\
vacuum
    &8.1
		&9.0
			&0.10
\\
\end{tabular}
\end{ruledtabular}
\end{table}

Due to the vacuum annealing, not only the electrical transport properties of the superconducting phase, but also those of the normal phase were affected.
As shown in the inset of Fig. \ref{Fig_rhoT}, $\rho_\mathrm{ab}$ of the vacuum-annealed \#RT shows metallic behavior while $\rho_{ab}$ of the O$_2$-annealed \#RT increases with decreasing temperatures.
Similar results caused by the reduction of excess iron have been observed in the O$_2$-annealed Fe$_{1+y}$Te$_{1-x}$Se$_x$
\cite{Liu_PRB80, Sun_PRB89}.
Furthermore, such metallic behavior has been observed in Fe$_{1+y}$Te$_{1-x}$Se$_x$ in homogeneous superconducting crystals
\cite{Okazaki_JPSJ81}.
Therefore, we conclude that excess iron is reduced in vacuum-annealed Fe$_{1+y}$Te$_{1-x}$S$_x$.

\begin{figure}[t]
\begin{center}
\includegraphics[clip, width=0.5\textwidth, bb=0 0 450 550]{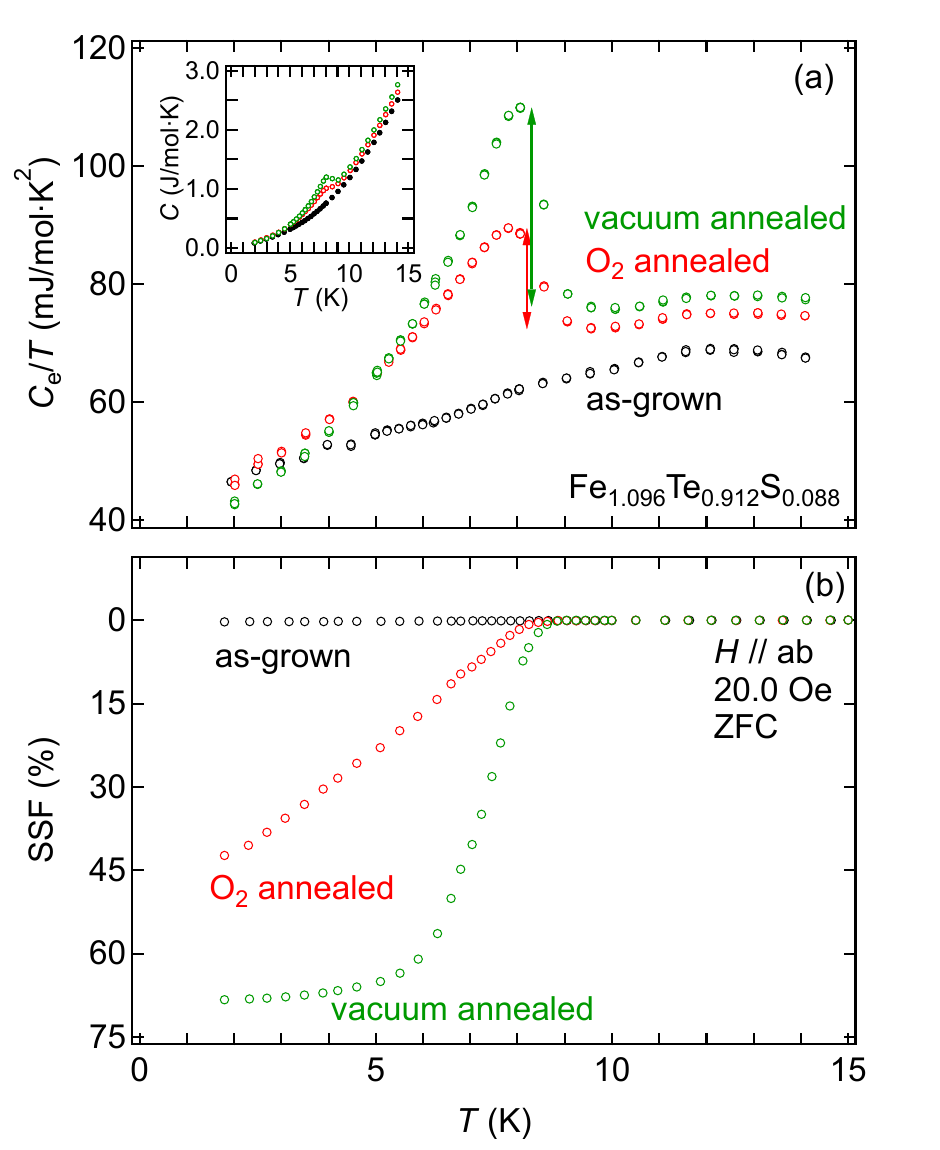}
\end{center}
\caption{
(a) Temperature dependence of the electronic specific heat divided by temperature, $C_\mathrm{e}/T$.
The vertical arrows indicate the specific heat-jump $\mathit{\Delta}C_\mathrm{e}$.
The inset in panel (a) shows the specific heat $C = C_\mathrm{e} + C_\mathrm{ph}$.
(b) Temperature dependence of the SSF of as-grown, O$_2$-annealed, and vacuum-annealed Fe$_{1.096}$Te$_{0.912}$S$_{0.088}$ (\#CT).
A magnetic field of 20.0 Oe was applied along the $ab$-plane.
}
\label{Fig_SpecificHeat}
\end{figure}

\begin{table*}[t]
\caption{
$\gamma_\mathrm{N}$ and $\beta$ obtained by the fit of specific heat $C$ with $C = \gamma_\mathrm{N} T + \beta T^3$, normalized specific heat-jump $\mathit{\Delta} C_\mathrm{e} / \gamma_\mathrm{J} T_\mathrm{c}$ at $T_\mathrm{c}$, and calculated Debye temperature $\mathit{\Theta}_\mathrm{D}$ of Fe$_{1.096}$Te$_{0.912}$S$_{0.088}$ (\#CT).
We also listed $T_\mathrm{c}^\mathrm{onset}$, volume $V$, diamagnetic factor $N_\mathrm{d}$, and SSF at 1.8 K.
}
\begin{ruledtabular}
\label{table_Debye}
\begin{tabular}{ccccccccc}
process
    &$\gamma_\mathrm{N}$ (mJ/mol$\cdot$K$^2$)
	   &$\beta$ (mJ/mol$\cdot$K$^4$)
        &$\mathit{\Delta} C_\mathrm{e} / \gamma_\mathrm{J} T_\mathrm{c}$
            &$\mathit{\Theta}_\mathrm{D}$ (K)
                &$T_\mathrm{c}^\mathrm{onset}$ (K)
                    &$V$ ($10^{-3}$ cm$^3$)
                        &$10^2 \times N_\mathrm{d}$
                            & SSF (1.8 K) (\%)
\\
\hline
as-grown
	&67.1  
		& 0.649   
			& 
                &144  
                    &
                        &1.08
                            &7.13
                                &
\\
O$_2$
    &73.7   
		& 0.568   
			& 0.546   
                &151  
                    &8.2
                        &0.51
                            &9.46
                                & 42.3
\\
vacuum
	&76.8  
		& 0.596   
            & 1.06   
			     &   148  
                & 8.3
                     &0.54
                            &10.5
                                &62.3
\\
\end{tabular}
\end{ruledtabular}
\end{table*}

For further investigation of the improvement in the superconducting volume fraction due to the multiple processes, we measured the specific heat of Fe$_{1.096}$Te$_{0.912}$S$_{0.088}$ (\#CT).
Figure \ref{Fig_SpecificHeat}(a) shows the temperature dependence of the electronic specific heat divided by the temperature, $C_\mathrm{e}/T$.
Here, $C_\mathrm{e}$ was calculated by subtracting the phonon specific heat $C_\mathrm{ph}$ from the total specific heat $C$ in the inset of Fig. \ref{Fig_SpecificHeat}(a).
$C_\mathrm{ph}$ is obtained by fitting $C$ above 9 K using $C_\mathrm{e} = \gamma_\mathrm{N} T$ and $C_\mathrm{ph} = \beta T^3$.
We list the obtained values of $\gamma_\mathrm{N}$ and $\beta$ in Table \ref{table_Debye}.
We observed a peak structure due to the superconducting transition in $C_\mathrm{e}/T$ in the O$_2$- and vacuum-annealed \#CT around $T_\mathrm{c}$ where the SSF in Fig. \ref{Fig_SpecificHeat}(b) shows diamagnetism.
The peak magnitude of $C_\mathrm{e}/T$ is larger for the vacuum-annealed \#CT compared to the O$_2$-annealed one.
Therefore, we conclude that the vacuum annealing in addition to the O$_2$ annealing is a good method to improve the superconducting volume fraction.

Assuming entropy balance and the exponential decay of the $C_\mathrm{e}/T$ below $T_\mathrm{c}$
\cite{Nishizaki_JPSJ67}, 
we can also estimate the SSF by the normalized specific-heat jump, denoted as 
$\mathit{\Delta} C_\mathrm{e} / \gamma_\mathrm{J}T_\mathrm{c}$.
Here, $\gamma_\mathrm{J} = \gamma_\mathrm{N} - \gamma_0 = 42.6$ (41.3) mJ/mol$\cdot$K$^2$ is the superconducting contribution for the electronic specific heat of O$_2$-annealed (vacuum-annealed) sample, and $\gamma_0$ is the residual coefficient at $T = 0$.
Based on $T_\mathrm{c}$ listed in Table \ref{table_Debye}, $\gamma_\mathrm{J}$, and the BCS prediction of
$\mathit{\Delta} C_\mathrm{e} / \gamma_\mathrm{J}T_\mathrm{c} = 1.43$
\cite{Tinkham_Text},
we can estimate the SSF of the O$_2$-annealed and vacuum-annealed samples to be 38.2\% and 74.1\%, respectively.
The result is qualitatively consistent with the increase in SSF due to the vacuum annealing estimated from the magnetization measurements (see Table \ref{table_Debye}).

We also calculated the Debye temperature, given by 
$\mathit{\Theta}_\mathrm{D} = \left[ \left( 12 \pi^4 N k_\mathrm{B} \right) / \left( 5 \beta \right) \right]^{1/3 }$, of each processed sample in Table \ref{table_Debye}.
Here, $N$ is the number of formula units, and $k_\mathrm{B}$ is the Boltzmann constant.
These values are comparable to previous studies on Fe$_{1+y}$Te$_{1-x}$S$_x$
\cite{Yamazaki_JPSJ85, Chen_PRB79}.

As discussed above, we observed the increase in SSF, the narrowing of $\mathit{\Delta}T_\mathrm{c}/T_\mathrm{c}$ in $\rho_{ab}$, and the increase in peak magnitude of $C_\mathrm{e}/T$ with the application of the HCl etching and vacuum annealing to the O$_2$-annealed Fe$_{1+y}$Te$_{1-x}$S$_x$.

\subsection{
\label{results_MOI}
Magneto-optical imaging
}

\begin{figure*}[t]
\begin{center}
\includegraphics[clip, width=1.0\textwidth, bb=0 0 600 280]{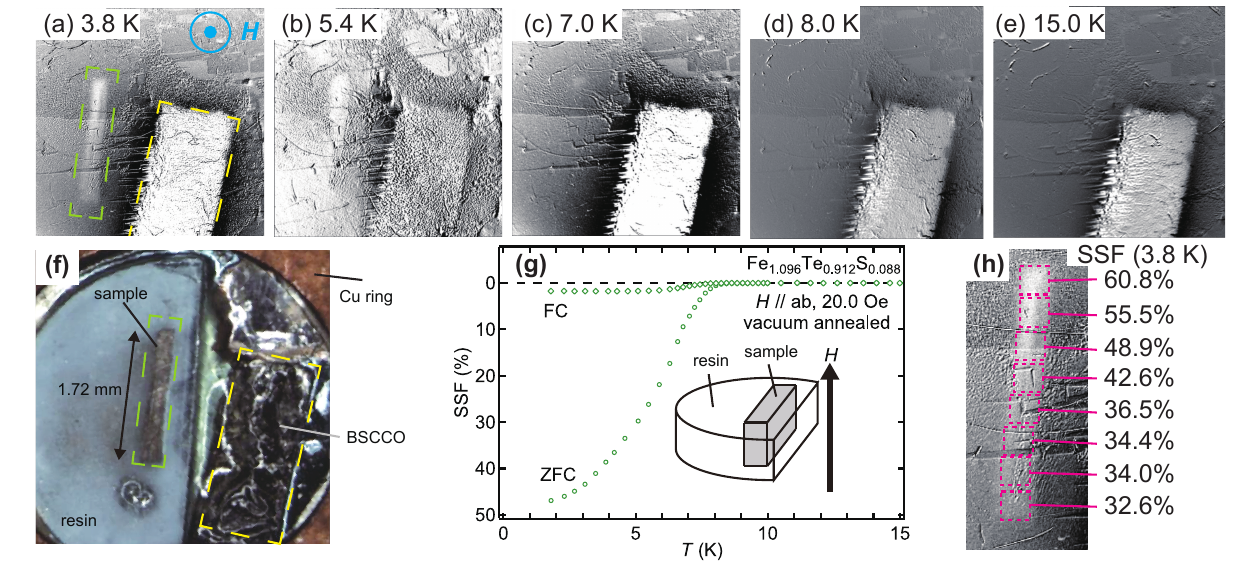}
\end{center}
\caption{
MO images at (a) 3.8 K, (b) 5.4 K, (c) 7.0 K, (d) 8.0 K, and (e) 15.0 K of the vacuum-annealed Fe$_{1.096}$Te$_{0.912}$S$_{0.088}$ (\#MO) under an external magnetic field of 12.6 Oe.
The MO images were taken during the ZFC process.
The contrast of each picture is calibrated for clarity.
The magnetic field direction is indicated in the panel (a).
(f) Backside picture of the MO imaging setup.
The green (yellow) broken boxes in panels (a) and (f) indicate the position of the sample \#MO (BSCCO). 
(g) Temperature dependence of the SSF of the vacuum-annealed \#MO.
The inset exhibits the schematic view of the sample and the magnetic field direction.
(h) SSF at 3.8 K calculated from the MO image data at several regions.
}
\label{Fig_MOI_TempDep}
\end{figure*}

To confirm bulk superconductivity in the vacuum-annealed samples, we focused on MO imaging measurements.
Figures \ref{Fig_MOI_TempDep}(a) - \ref{Fig_MOI_TempDep}(e) show the MO images of the vacuum-annealed \#MO at several temperatures.
Here, these images were taken at an external magnetic field of 12.6 Oe during the ZFC process.
Each MO image is obtained by subtracting the data at 0 Oe from that at 12.6 Oe.
We also exhibit the backside image of the MO imaging setup in Fig. \ref{Fig_MOI_TempDep}(f).
To identify the position of the sample \#MO, we also placed a piece of the high-$T_\mathrm{c}$ superconductor Bi$_2$Sr$_2$CaCu$_2$O$_{8+\delta}$ (BSCCO) with $T_\mathrm{c} \sim 110$ K.
At 3.8 K, well below $T_\mathrm{c}$, we observed diamagnetic signals of the vacuum-annealed \#MO.
Comparing this MO image with the photograph of the \#MO in Fig. \ref{Fig_MOI_TempDep}(f), we can confirm that the superconducting region well matches the same as the cross-sectional area of the sample.
Therefore, we conclude that the superconductivity of the vacuum-annealed \#MO is bulk.
At 5.4 and 7.0 K, the diamagnetic signals show a weakening and the superconducting regions exhibit narrowing.
At 8.0 K just below $T_\mathrm{c}^\mathrm{onset} = 8.6$ K of \#MO (see Fig. \ref{Fig_MOI_TempDep}(g) and Table \ref{table_MOI}), the diamagnetic signals are hardly visible.
The absence of diamagnetic signals at 15.0 K, which is above $T_\mathrm{c}$, is a natural result.
This temperature dependence of the diamagnetic signals in the MO images is qualitatively consistent with the SSF of the sample \#MO  (see Fig. \ref{Fig_MOI_TempDep}(g)).

\begin{table}[t]
\caption{
Information of the sample for MO imaging measurements in Fe$_{1.096}$Te$_{0.912}$S$_{0.088}$.
The values of SSF for the ZFC process are listed.
}
\begin{ruledtabular}
\label{table_MOI}
\begin{tabular}{cccc}
$T_\mathrm{c}^\mathrm{onset}$
	& $V$ ($10^{-3} $cm$^3$)
		& $10^2 \times N_\mathrm{d}$
			& SSF(1.8 K) (\%)
\\
\hline
8.6
	& 0.12
		& 4.54
			& 47.3
\end{tabular}
\end{ruledtabular}
\end{table}

While bulk superconductivity is indicated by the MO images at 3.8 K, the value of SSF varies depending on the position of the sample along the in-plane direction.
Based on the gray values of the MO images at 3.8 K, we estimate the strength of the diamagnetic fields in several regions as indicated in Fig. \ref{Fig_MOI_TempDep}(h).
The highest SSF of 60.8\% is obtained at the upper end of the sample, and this value decreases from the upper end to the lower end.
This result suggests that the superconductivity of the vacuum-annealed Fe$_{1+y}$Te$_{1-x}$S$_x$ is not fully homogenized.
Thus, the existence of further optimized annealing conditions can be pursued.
It is noted that the average value of eight SSFs, which is estimated to be 43\%, is consistent with the SSF determined by the magnetization measurements in Fig. \ref{Fig_MOI_TempDep}(g).

\begin{figure}[h]
\begin{center}
\includegraphics[clip, width=0.5\textwidth, bb=0 0 520 460]{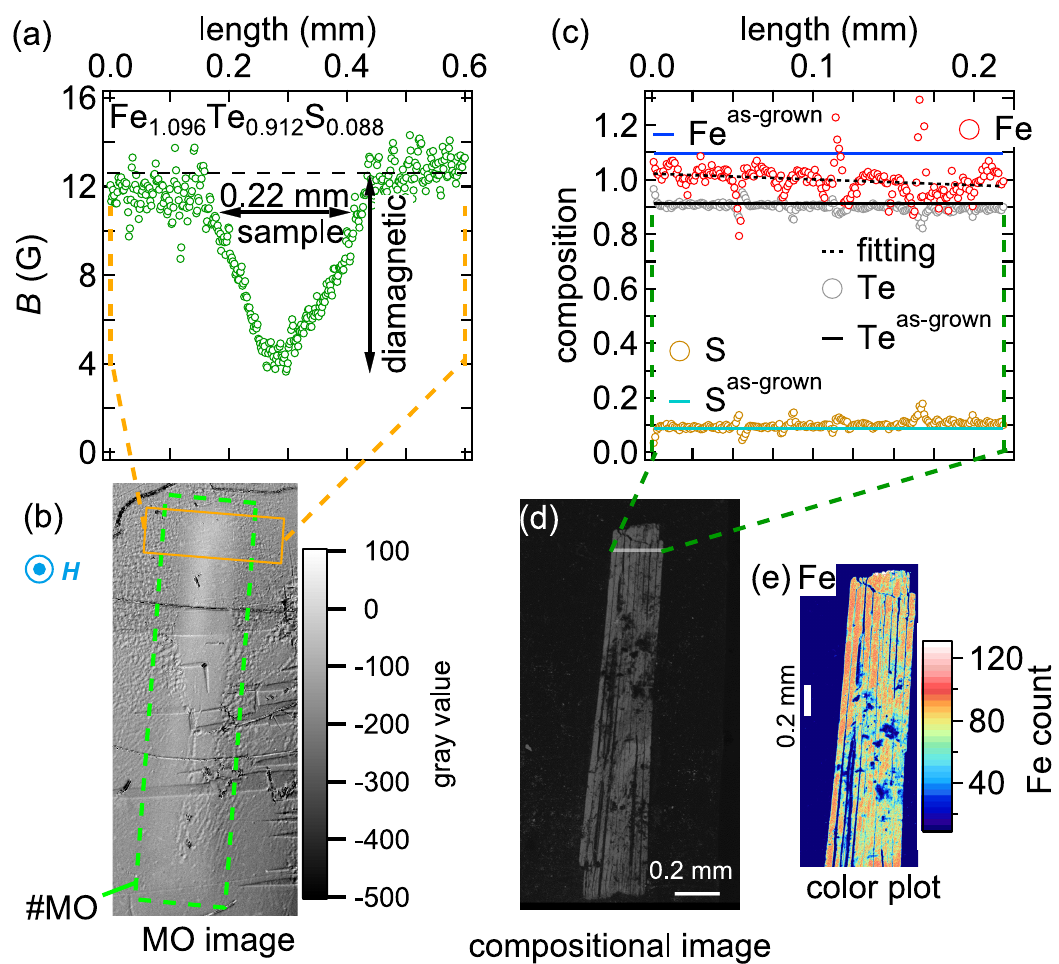}
\end{center}
\caption{
(a) Calculated distribution of magnetic fields and (b) MO image of vacuum-annealed Fe$_{1.096}$Te$_{0.912}$S$_{0.088}$ (\#MO) in the superconducting state at 3.8 K for ZFC process.
The distribution is calculated from the yellow bold line in the MO image.
The orange broken lines indicate the edge of the distribution.
The green broken box indicates a guide for the position of \#MO.
The black broken line indicates the value of the applied external magnetic field 12.6 Oe.
The magnetic field direction is also indicated. 
(c) Composition distributions and (d) compositional image of the vacuum-annealed \#MO determined by WDS. 
The distributions are measured along the white bold line in the compositional image of \#MO.
The blue, black, and light blue solid lines indicate the Fe, Te, and S compositions of the as-grown state of \#MO, respectively.
The black broken line shows the linear fit of the Fe composition of the vacuum-annealed \#MO.
(e) Color plot of the x-ray count from Fe atoms existing near the surface of the vacuum-annealed \#MO.
}
\label{Fig_MOI_3p8K}
\end{figure}

In contrast to the magnetic field distribution along the in-plane direction, a relatively symmetric distribution indicating homogeneous superconductivity along the out-of-plane direction was obtained.
Figure \ref{Fig_MOI_3p8K}(a) shows the position dependence of the magnetic field distribution under an external magnetic field of 12.6 Oe along the $c$-axis of \#MO.
The yellow bold line in Fig. \ref{Fig_MOI_TempDep}(g) indicates the analysis region, where the SSF is estimated to be 60.8\%.
In the sample, the value of the magnetic field distribution obtained from the MO image is lower than the external magnetic field of 12.6 Oe.
While the diamagnetic signals show weakening near the edge of the sample due to the superconducting shielding currents under the external magnetic field, the diamagnetic response is the strongest near the center of the sample.
These results demonstrate that the superconductivity of the vacuum-annealed \#MO is bulk and homogeneous along the out-of-plane direction.

As described above, the MO measurements show that the superconductivity of the vacuum-annealed Fe$_{1.096}$Te$_{0.912}$S$_{0.088}$ is bulk.

\subsection{
\label{EPMA}
Compositional analysis
}

Based on the previous studies
\cite{Okazaki_JPSJ81},
the removal of excess iron is crucial for achieving superconductivity.
To investigate the relationship between the realization of such bulk superconductivity and the residual excess iron, we performed the directional quantitative analysis of the Fe, Te, and S compositions in the MO-imaged region after the MO measurements (see Fig. \ref{Fig_MOI_3p8K}(c)).
In addition, the Fe, Te, and S compositions of the as-grown \#MO are shown. 
We also show the compositional image, which displays the regions of higher material density in white, of the cross-sectional area measured by WDS in Fig. \ref{Fig_MOI_3p8K}(d).
Due to the cleavage of the sample after the MO measurements, several peak and dip structures are observed in the distribution of compositions.
The Te and S compositions of the vacuum-annealed state are almost the same as those of the as-grown state.
On the other hand, based on the linear fit of the composition distribution of Fe within the range of 0.22 mm shown in Fig. \ref{Fig_MOI_3p8K}(c), denoted as $(1.022 \pm 0.007) - (0.209 \pm 0.059) \times [\mathrm{length} / \mathrm{mm}]$, we can conclude that the Fe composition of the vacuum-annealed state decreases to below that of the as-grown value of 1.096.
e also attempted to measure the Fe composition at the lower edge of the sample, however, we were unable to obtain a clear result due to the cleavage (see  Figs. \ref{Fig_MOI_3p8K}(d) and \ref{Fig_MOI_3p8K}(e)).
Although no results of the directional analysis were obtained in other regions, we performed a two-dimensional qualitative analysis of the Fe composition shown in Fig. \ref{Fig_MOI_3p8K}(e).
This result suggests that the composition of the excess iron in \#MO is less than $1+y = 1.096$. 
In addition, the Fe composition of the bottom of the sample is lower than that of the top (see Fig. \ref{Fig_MOI_3p8K}(e)).
This iron distribution is consistent with the MO images at several temperatures in Figs. \ref{Fig_MOI_TempDep}(a) - \ref{Fig_MOI_TempDep}(e), demonstrating that the superconducting transition temperature $T_\mathrm{c}$ of the bottom of the sample is lower than the $T_\mathrm{c}$ of its top.
Based on these results, therefore, we conclude that the vacuum annealing after the HCl etching can remove excess iron for the O$_2$-annealed Fe$_{1+y}$Te$_{1-x}$S$_x$.
It is noted that the effectiveness of vacuum annealing in removing excess iron is also observed in another vacuum-annealed sample (see Appendix \ref{Appendix_COMP}).

While the compositional analysis can confirm the removal of excess iron, its homogeneity depends on the shape of the sample.
The directional analysis in Fig. \ref{Fig_MOI_3p8K}(c) demonstrates that the distribution of excess iron is uniform along the out-of-plane direction.
On the other hand, the color plot in Fig. \ref{Fig_MOI_3p8K}(e) shows that the amount of excess iron is lower at the bottom of the sample than at the top.
In the present studies, we prepared thin samples in the $c$-axis direction.
Considering these results, we deduce that the excess iron tends to be more easily removed along the thinner direction of the sample.
This tendency is consistent with the previous studies that investigated the thickness dependence of the removal of excess iron by the vacuum-annealing process
\cite{Dong_PRM3}
and the origin of the non-homogeneous SSF shown in Fig. \ref{Fig_MOI_TempDep}(h).


\section{
\label{discussion}
Discussion
}

\begin{figure*}[htbp]
\begin{center}
\includegraphics[clip, width=0.8\textwidth, bb=0 0 580 190]{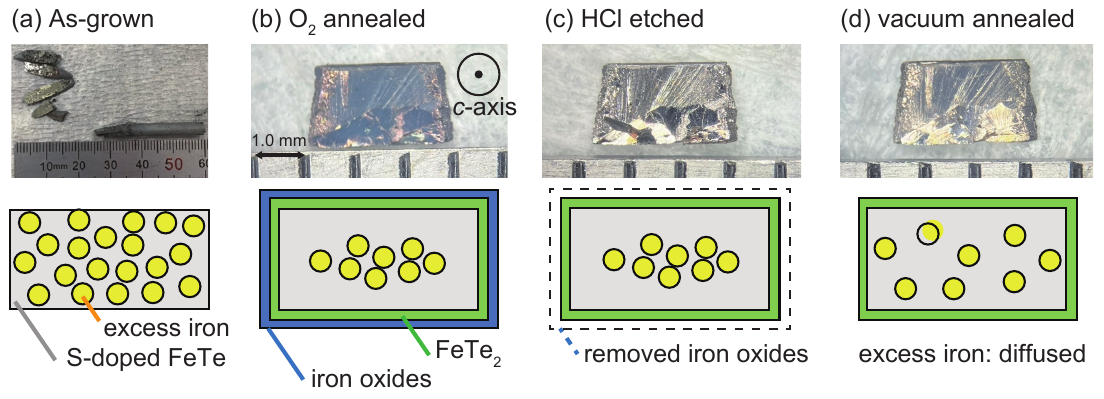}
\end{center}
\caption{
Photographs and schematic views of a sample in (a) as-grown, (b) O$_2$ annealed, (c) HCl etched, and (d) vacuum-annealed states.
}
\label{Fig_SchematicViewOfSample}
\end{figure*}

We demonstrated the realization of bulk superconductivity in the processed Fe$_{1+y}$Te$_{1-x}$S$_x$ by physical property and MO measurements.
Furthermore, the compositional analysis by WDS showed that the excess iron was homogeneously removed from the samples along the out-of-plane direction.
Based on the above results, we discuss the relationship between superconductivity and excess iron.

Figure \ref{Fig_SchematicViewOfSample} shows the photographs and schematic views of the processed Fe$_{1+y}$Te$_{1-x}$S$_x$.
In the as-grown state, excess iron is distributed throughout the sample (see Fig. \ref{Fig_SchematicViewOfSample}(a)).
Although a small region in as-grown Fe$_{1+y}$Te$_{1-x}$S$_x$ shows superconductivity
\cite{Okazaki_JPSJ81},
excess iron prevents its emergence.
Thus, the superconducting responses of the as-grown sample were hardly observed in the magnetization and specific heat measurements.
Applying O$_2$-annealing to an as-grown sample allows excess iron to be removed and iron oxides to form near the surface of the sample.
However, FeTe$_2$ can also be formed simultaneously on the sample surface (see Fig. \ref{Fig_SchematicViewOfSample}).
The gradual decrease of the magnetization and resistivity below the onset of the superconducting transition temperature $T_\mathrm{c}^\mathrm{onset}$ indicates that the superconductivity is not homogeneous even in the best-optimized O$_2$-annealing condition in the present study.
The iron oxides formed on the surface were removed by the HCl etching process (see Fig. \ref{Fig_SchematicViewOfSample}), but FeTe$_2$ remains.
The magnetization of the HCl-etched \#MT still showed a gradual decrease below $T_\mathrm{c}^\mathrm{onset}$, and the value of SSF is comparable to that of the O$_2$-annealed state, suggesting that the inhomogeneous distribution of excess iron in the sample remains.
Applying the vacuum-annealing process to the HCl-etched sample resulted in a sharpening of the diamagnetic response and a decrease in resistance below $T_\mathrm{c}^\mathrm{onset}$.
These results are consistent with the homogeneous superconductivity of a small piece of Fe$_{1+y}$Te$_{1-x}$S$_x$
\cite{Okazaki_JPSJ81}.
An increase in the peak height of the specific heat around the superconducting transition was also observed, indicating the realization of bulk superconductivity.
The visualized diamagnetic response in the cross-sectional area of vacuum-annealed \#MO also confirmed bulk superconductivity.
Therefore, we concluded that the diffusion of excess iron in the sample was caused by the vacuum-annealing process (see Fig. \ref{Fig_SchematicViewOfSample}).
The decrease in the peak intensity of FeTe$_2$ may also imply the re-diffusion of Fe atoms within the crystal.
The uniform removal of excess iron along the thinner direction can be confirmed by the compositional analysis by WDS.

Our experimental results suggest that FeTe$_2$ is formed on the surface of the processed sample, however, further verification can be needed.
Since the compositional distribution of Fe, Te, and S atoms does not show a significant decrease in Fe atoms near the sample surface (see Figs. \ref{Fig_MOI_3p8K}(c) and \ref{Fig_EPMA}(a)), FeTe$_2$ can only be formed to a depth of a few microns or less from the sample surface considering the spatial resolution of EPMA.
We desire the actual thickness of FeTe$_2$ formed on the sample surface in each processed state for further optimization of the superconductivity in S-substituted Fe$_{1+y}$Te and related iron-based superconductors.


\section{
\label{conclusion}
Conclusion}

We investigated the effect of multiple chemical processes in S-substituted Fe$_{1+y}$Te to obtain bulk superconductivity.
We focused on the vacuum-annealing for the O$_2$-annealed sample after the HCl etching process.
For efficient annealing, we investigated optimized conditions for oxygen annealing.
The effectiveness of the vacuum annealing was confirmed by the increase in the superconducting shielding fraction, the rapid decrease in magnetization and resistivity below the onset of the superconducting transition temperature, and the increase in the peak structure of the specific heat around the superconducting transition.
In addition to these physical property measurements, we observed the diamagnetic response in the magneto-optical imaging, indicating that the superconductivity induced by the vacuum-annealed process was bulk.
The removal and actual composition of excess iron was confirmed by WDS. 
These results support the importance of the removal of excess iron for the realization of the bulk superconductivity in Fe$_{1+y}$Te$_{1-x}$S$_x$.

\section*{Acknowledgment}
We appreciate Y. Soma and K. Naito for their experimental assistance.
Our magnetic property measurements were carried out using the facilities of the Materials Design and Characterization Laboratory by the joint research in the Institute for Solid State Physics, The University of Tokyo (No.202204-MCBXU-0045, 202205-MCBXG-0054, 202212-MCBXG-0012, 202305-MCBXG-0059, 202311-MCBXG-0031).

This work was partly supported by JSPS transformative research areas (A), section (II) (JP 23H04862).

\section*{Author contributions}
R. Kogure conceived and designed the experiments
R. Kogure also performed the sample preparations, annealings, physical property measurements, and analysis.
R. Kogure, T. Ota, and Y. Kinoshita measured the MO images.
S. Hakamada carried out experiments to respond to the Referee's comments.
M. Tokunaga supervised the conduct of the MO imaging measurements.
R. Kurihara drafted the original manuscript.
R. Kurihara and H. Yaguchi supervised the conduct of this study.
All authors checked the original manuscript and contributed to its revision.
All authors have agreed to submit the final version of the manuscript.

Most of this work is based on a thesis submitted to partially fulfill the requirement for the master's degree in Science of Ryusuke Kogure of the Graduate School of Science and Technology, Tokyo University of Science (FY2023).


\appendix

\section{
\label{Appendix_COMP}
Compositional analysis of another vacuum-annealed Fe$_{1+y}$Te$_{1-x}$S$_x$ sample
}

\begin{figure}[h]
\begin{center}
\includegraphics[clip, width=0.5\textwidth, bb=0 0 380 280]{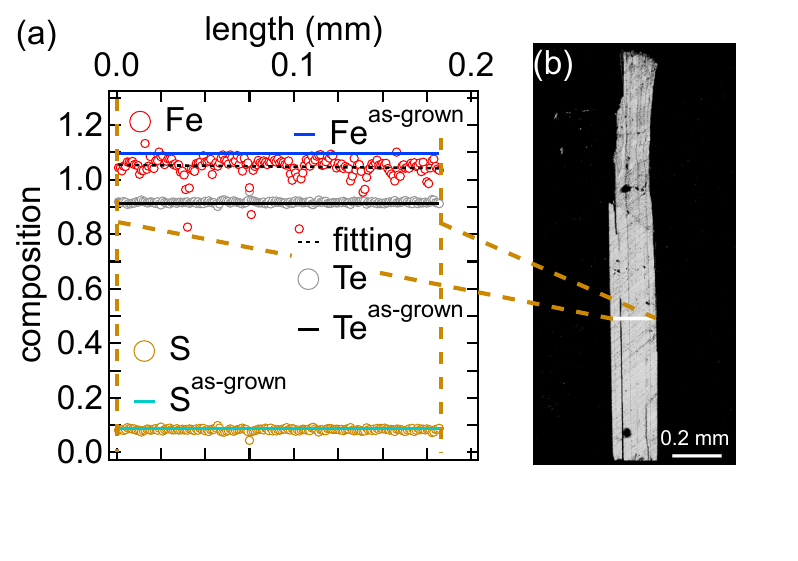}
\end{center}
\caption{
(a) Composition distributions and (b) compositional image of the vacuum-annealed Fe$_{1+y}$Te$_{1-x}$S$_x$ with $y = 0.096$ and $x = 0.088$ (\#COMP) determined by WDS.
The distributions are measured along the white bold line in the compositional image.
The blue, black, and light blue solid lines indicate the Fe, Te, and S composition of the as-grown state, respectively.
The black broken line shows the linear fit of the Fe composition of the vacuum-annealed \#COMP.
}
\label{Fig_EPMA}
\end{figure}

Figure \ref{Fig_EPMA}(a) shows the compositional distribution along the out-of-plane direction of Fe$_{1.096}$Te$_{0.912}$S$_{0.088}$ (\#COMP) by WDS in the cross-sectional area of the sample in \ref{Fig_EPMA}(b).
We also illustrate the Fe, Te, and S compositions of as-grown \#MO. 
Because of the cleavage of the sample after the MO measurements, several peak and dip structures are observed in the distribution of compositions.
Based on the linear fit of the compositional distribution of Fe within the range of 0.18 mm shown in Figure \ref{Fig_EPMA}(a), denoted as $(1.0548 \pm 0.00568) - (0.069968 \pm 0.0538) \times [\mathrm{length}/\mathrm{mm}]$, we can conclude that the Fe composition decreases to below 1.096.
Therefore, the vacuum annealing can reduce the excess iron from Fe$_{1+y}$Te$_{1-x}$S$_x$.

\bibliography{main.bib}

\end{document}